# Transformer-Based Hematological Malignancy Prediction from Peripheral Blood Smears in a Real-World Cohort

Muhammed Furkan Dasdelen[1,2,*], Ivan Kukuljan[1,*], Peter Lienemann[1,3,], Fatih Ozlugedik[1], Ario Sadafi[1], Matthias Hehr[4], Karsten Spiekermann[3], Christian Pohlkamp[5], and Carsten Marr[1,3,6]

[1]Institute of AI for Health, Helmholtz Zentrum München—German Research Center for Environmental Health, Neuherberg, Germany
[2]International School of Medicine, Istanbul Medipol University, Istanbul, Turkiye
[3]Department of Medicine III, Ludwig-Maximilian-University Hospital, Munich, Germany
[4]Dr. von Haunersches Kinderspital, Ludwig-Maximilians-University Munich, Munich, Germany
[5]Munich Leukemia Laboratory, Munich, Germany
[6]DKTK, German Cancer Consortium, Germany

* Shared first authorship
Correspondence: carsten.marr@helmholtz-munich.de

**Abstract**

Peripheral blood smears remain a cornerstone in the diagnosis of hematological neoplasms, offering rapid and valuable insights that inform subsequent diagnostic steps. However, since neoplastic transformations typically arise in the bone marrow, they may not manifest as detectable aberrations in peripheral blood, presenting a diagnostic challenge. In this paper, we introduce cAItomorph, an explainable transformer-based AI model, trained to classify hematological malignancies based on peripheral blood cytomorphology. Our data comprises peripheral blood single-cell images from 6115 patients with diagnoses confirmed by cytomorphology, cytogenetics, molecular genetics, and immunophenotyping from bone marrow samples, and 495 healthy controls, eight coarse classes. cAItomorph leverages the DinoBloom hematology foundation model and aggregates image encodings via a transformer-based architecture into a single vector. It achieves an overall accuracy of 0.72 in eight disease classification, with F1 scores of 0.76 for acute leukemia, 0.80 for myeloproliferative neoplasms and 0.94 for healthy cases. The overall accuracy increases to 0.87 in top-2 predictions. cAItomorph achieves high sensitivity for acute leukemia cases in external test sets. By analyzing attention heads, we demonstrate clinically relevant cell-level attentions in both internal and external test sets. Moreover, our model's calibrated prediction probabilities reduce the false discovery rate from 13.5% to 8.7% without missing any acute leukemia cases, thereby decreasing the number of unnecessary bone marrow aspirations based on peripheral blood smears. This study highlights the potential of AI-assisted diagnostics in hematological malignancies, illustrating how models trained on real-world data could enhance diagnostic accuracy and reduce invasive procedures.

**Key messages:**

- We assemble the first real-world dataset of patient blood smears for hematological malignancies.
- We introduce a state-of-the-art AI model for peripheral blood diagnostics based on a hematology foundation model.
- The model reaches an excellent performance for disease classes that are diagnosable from peripheral blood.
- The model can support human experts in estimating probabilities of hematological malignancies and help in guiding the downstream diagnostic tasks.

# 1. Introduction

Hematological malignancies represent a wide range of disease entities, most of which arise from dysfunctional proliferation and differentiation of hematopoietic stem and progenitor cells in the bone marrow. They account for 6.5% of all estimated cancer cases worldwide[1]. Some of them, like Acute Myeloid Leukemia (AML), still have a 5-year survival rate as low as 30-35%[2,3]. Diagnostic procedures for hematological cancers comprise cytomorphology, cytogenetics, immunophenotyping, and molecular genetics. The first step is assessing the differential cell counts and cellular morphological abnormalities in a peripheral blood smear analysis. The World Health Organization (WHO) classification[4] outlines specific criteria for hematologic conditions, demonstrating the critical role of peripheral blood smears in identifying characteristic cellular abnormalities. Unlike the painful bone marrow aspiration, a blood smear is fast, minimally invasive, and provides valuable information that guides the follow-up diagnostic pathway. However, conventional peripheral blood smear analysis involves labor-intensive manual examination of hundreds of cells under the microscope. While human investigators can identify diagnostic clues and certain cellular abnormalities in peripheral blood for diseases such as acute leukemias, myeloproliferative neoplasms, and chronic lymphocytic leukemia, they often cannot definitively confirm or rule out a hematological malignancy based solely on a peripheral blood smear. Moreover, manual examination suffers from intra- and inter-rater variability[5], may delay diagnosis and treatment, and requires a trained cytologist. The demand for trained experts is increasing in high-income countries, while there is a serious shortage in low-resource countries[6,7].

Deep learning algorithms offer significant benefits for image-based classification tasks. They can automate the analysis process, reducing the workload on healthcare professionals, and increasing the accuracy and speed of diagnosis[8–10]. They are particularly crucial for diseases like acute lymphoblastic leukemia (ALL) and acute myeloid leukemia (AML), where timely intervention is vital[4]. Previous efforts demonstrated that supervised deep learning algorithms can achieve excellent results in white blood cell classification and malignant cell identification[5,11–17]. Moreover, state-of-the-art digital pathology applications can detect different types of leukemia from bone marrow[18–20] and peripheral blood smears[18,21,22]. Most of these supervised deep-learning models require costly single-cell annotations. In our previous study, we thus employed a multiple instance learning algorithm[23] for AML subtype classification with patient-level labels[24], achieving an F1 score of 0.86±0.05 and eliminating the need for single-cell annotation. However, all previous studies were based on carefully curated datasets, specific to a particular disease class.

Real-world clinical data, in contrast, is heterogeneous, noisy, and contains a plethora of disease types. It is unclear how these algorithms perform in actual clinical settings that cytologists encounter daily. To investigate this question, we require a real-world dataset comprising diagnostic labels and digitized peripheral blood smear images, and a high-performance self-supervised learning model with integrated explainability.

In this study, we present the first real-world peripheral blood cytomorphology dataset encompassing the full spectrum of hematological diseases, with patient fractions reflecting their population-level distribution. Using the DinoBloom hematology foundation model[25], we train a transformer-based AI model to classify eight coarse categories of hematological malignancies and to predict hemoglobin values from peripheral blood smear single-cell images, using only patient-level labels. Our model, cAItomorph, achieves high accuracy in distinguishing acute leukemia, myeloproliferative neoplasms, and healthy conditions, and demonstrates surprisingly strong performance on plasma cell neoplasms and MDS. Moreover, the predicted hemoglobin values show a high correlation with ground-truth measurements. We show that cAItomorph's output probabilities are well calibrated and can be used to identify the presence of malignancies. The model's performance is validated on external datasets and further tested on diseases unseen during training. Finally, cell-level attention maps derived from the transformer architecture provide interpretability and align with clinical decision-making.

## 2. Methods

### 2.1. Data

The real-world dataset comprises 6115 patients first time diagnosed in the Munich Leukemia Laboratory (MLL) during the years 2021-2022, along with 495 healthy donors (Figure 1). Peripheral blood and bone marrow samples were collected from every patient (Figure 1A). Diagnoses were made based on bone marrow cytomorphology, immunophenotyping, and cyto- and molecular genetics, as defined by the WHO guidelines[4]. These diagnoses served as the ground truth labels for our patients.

Single white blood cell images were obtained as described previously[24]. Briefly, Wright-Giemsa stained peripheral blood smears were initially scanned using a 10x objective, producing an overview image. The Metasystems Metafer software then detected high quality single leukocyte images after a segmentation threshold and logarithmic color adjustment were applied. The largest possible number of leukocytes with sufficient quality were positioned in each image and scanned with a 40x objective. Images were stored in TIFF format with 144x144 pixels. Additionally, basic information about the patients (age, sex, and blood counts) is available.

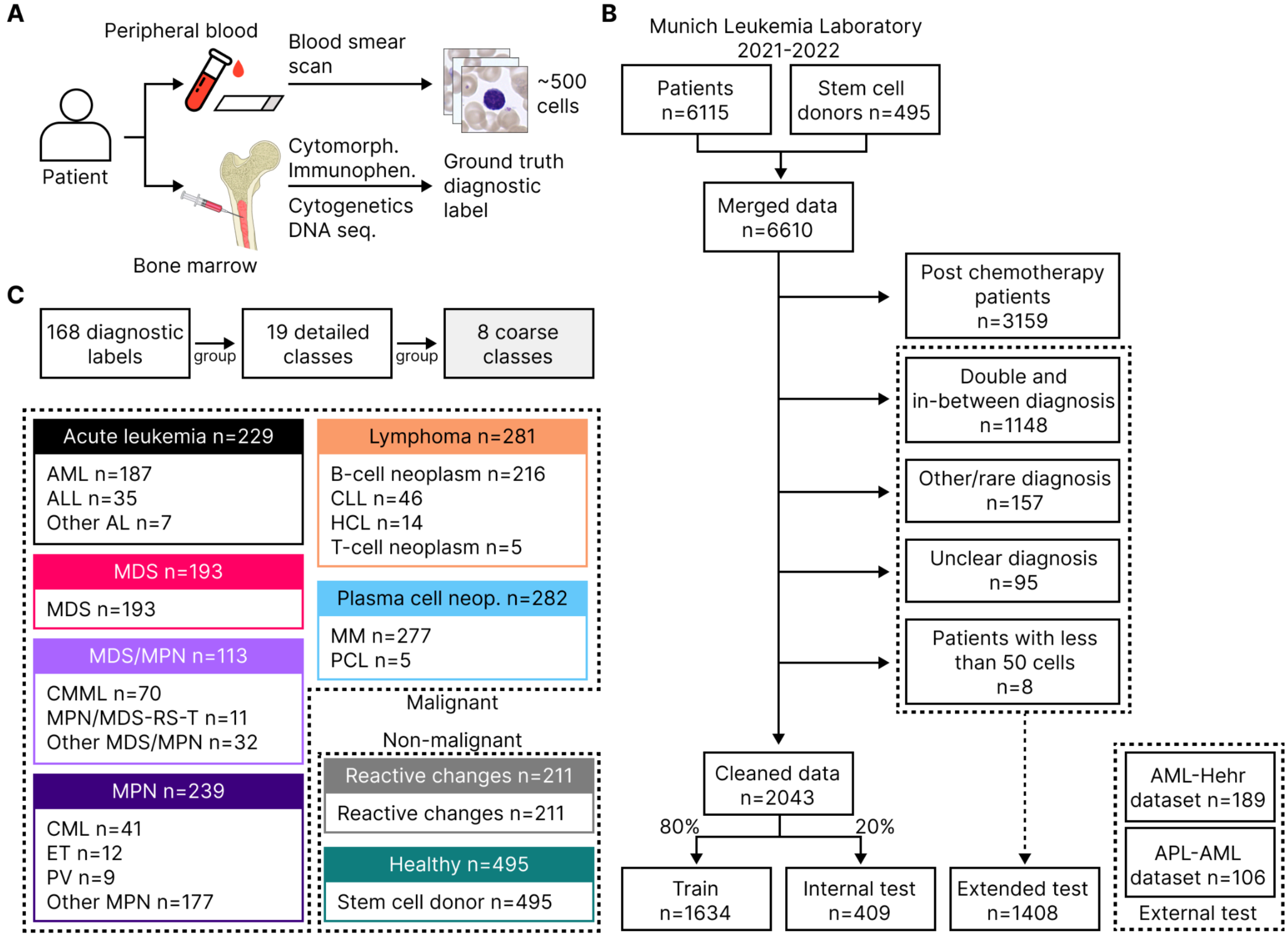

**Figure 1. Our real-world leukemia laboratory dataset contains over 3.2 million blood cell images from blood smears of 6115 patients and 495 controls and provides labels for 19 detailed and 8 coarse hematological disease classes. (A)** From all patients included in the cohort, both a peripheral blood sample and a bone marrow sample were collected. We used up to 500 blood cell images per patient for AI-based diagnosis. Cytomorphology, DNA sequencing, immunophenotyping and cytogenetics from bone marrow were used to determine the ground truth diagnosis. **(B)** During the data cleaning process, we removed follow-up patients since we focussed on initial diagnoses. We also discarded cases with double and in-between diagnoses, unclear diagnoses, and other diagnoses. These were

later reintroduced in the extended test set. **(C)** 168 diagnostic labels were categorized into 18 detailed classes, which were then further consolidated into broader 8 coarse classes.

### 2.2. Data cleaning and diagnostic label grouping

We cleaned the dataset by excluding patients with post-chemotherapy follow-ups (n=3159), unclear (n=95) or double diagnoses (n=1148), patients with few cells (n=8) and rare or undetectable conditions (n=143) (Figure 1B). Parts of this excluded data were reintroduced later in the extended test set (see Results). After cleaning, our dataset comprised 2043 patients with a total of 996,087 single-cell images. The final cleaned data includes 478±55 single white blood cell images per patient.

The dataset contains 168 different diagnosis labels, some common and some rare. We grouped the labels into 19 detailed classes, such as "AML" (including subtypes), "B-cell neoplasm", or "CMML". We then grouped diseases into 8 coarse classes, namely "Acute leukemia", "Lymphoma", "MDS", "MDS/MPN", “MPN”, “Plasma cell neoplasm” , “Reactive changes” and “Healthy” (Figure 1C).

We reserved 20% of the data (409 cases) for testing by stratifying according to detailed classes.

### 2.3. External datasets

We test our model on two external dataset. AML-Hehr[24] includes 129 patients diagnosed with four most prevalent AML subtypes with defining genetic abnormalities and 60 healthy stem cell donors. The dataset consists of 430±107 single white blood cell images per patient. APL-AML[26] has 106 patients, all diagnosed with acute leukemia and categorized into two: acute promyelocytic leukemia (APL) and other AML subtypes.

### 2.4. AI architecture and training

The task involves assessing around 500 single leukocyte images per patient, and determining the correct patient level diagnosis. This problem is categorized as weakly supervised learning and has been previously applied in hematology.[23,24,27,28] Our models consist of three steps: (i) Latent space encoding of single-cell images, (ii) feature vector aggregation, and (iii) classification (Figure 2A). The goal of encoding is to compress 144x144 pixel single-cell images into a smaller feature vector. In the second step, we aggregate the feature vectors from all white blood cells into one single vector. This vector is then utilized by a classifier to predict the diagnosis.

For the encoding part, we use DinoBloom hematology foundation model (ViT-B)[25]. For the aggregation, we use a modified vision transformer architecture (ViT)[29,30] with 10 transformer layers, 8 attention heads of dimension 64 resulting in an embedding dimension of 512, and a hidden dimension of 2,048. We ablate the number of layers and token pooling in comparison to other multiple instance learning aggregators (Supplementary table 1, 2). The MLP (multilayer perceptron) classifier consists of 2 layered neural networks with 128 hidden dimensions.

The model is trained with AdamW optimizer, weight decay of 0.01, learning rate of 5e-5 for 150 epochs with early stopping based on validation loss. We use cross entropy loss for disease classification and mean square error for hemoglobin prediction.

### 2.4. Model ensembling

For model development, we conduct 5-fold cross-validation (Figure 2A). The softmax function is applied to the output logits to obtain class probabilities. During testing, the five trained models are treated as independent learners, and their predicted probabilities are averaged to produce ensemble outputs. To derive the overall malignancy probability, we compute the sum of the probabilities corresponding to Acute leukemia, Lymphoma, MDS, MDS/MPN, MPN, and Plasma cell neoplasm classes.

### 2.5. Explainability

In medical AI applications, explainability is crucial to ensure that an algorithm bases its decisions on the correct details within the data. We obtain cell level attentions from transformer heads using Attention Rollout method[31]. This method provides insights into how each image contributes to the diagnostic decision.

### 2.6. Evaluation metrics

For model performance evaluation, we calculate sensitivity, precision, and F1 score. Sensitivity is the ratio of true positive cases among all actual positive cases. Precision is the ratio of true positive cases among all predicted positive cases. The F1 score is the harmonic mean of sensitivity and precision. We also measure false discovery rate, which is calculated as $1 - precision$, representing the proportion of false positive predictions among all positive predictions made by the model.

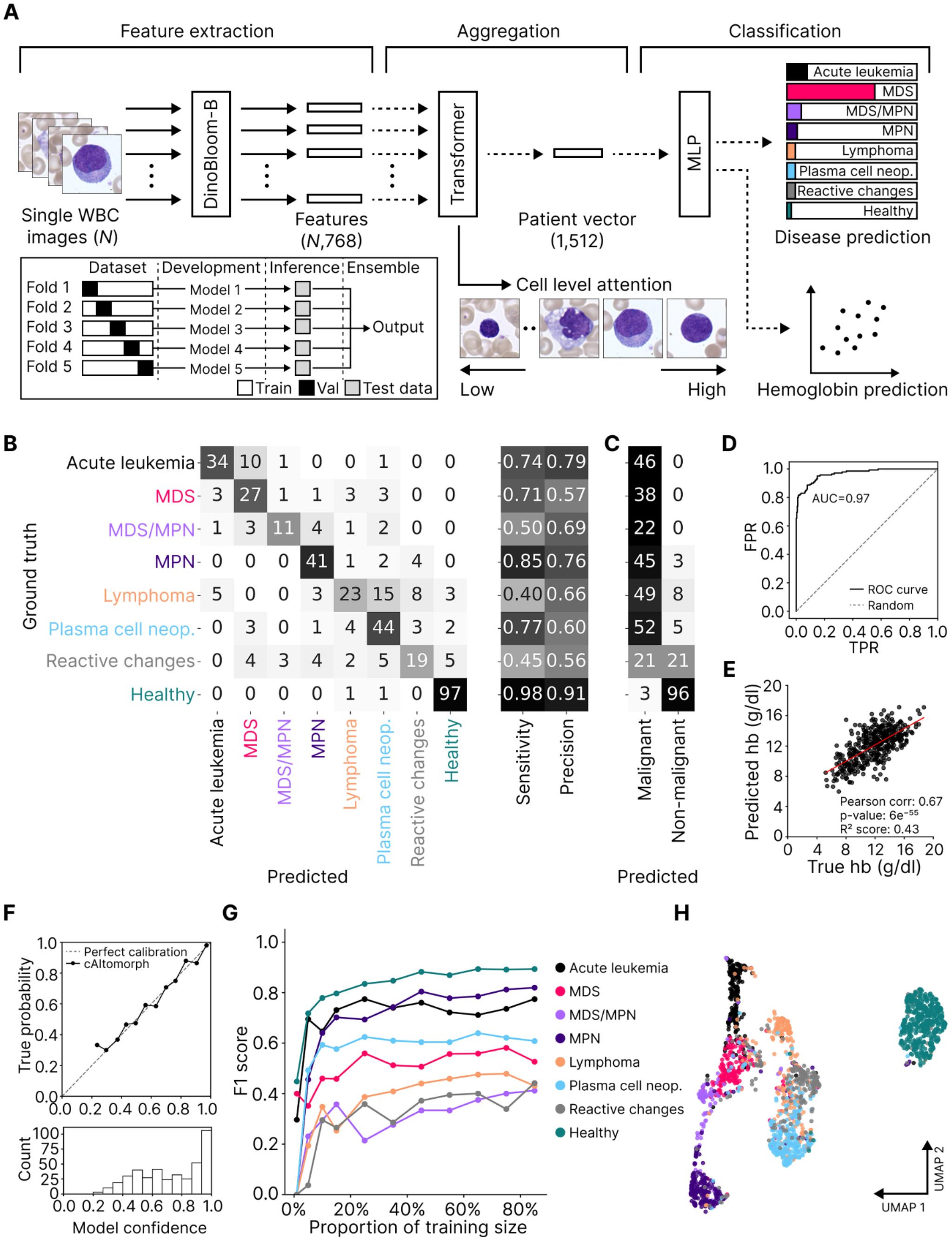

**Figure 2. cAItomorph predicts hematological malignancies and hemoglobin value from peripheral blood images. (A)** Model architecture: In the feature extracting stage, the state-of-the-art DinoBloom foundation model provides a latent space encoding of the single-cell images. During aggregation, we combine the single-cell encodings into a single latent space vector using a transformer. For classification, we employ a multi-layer perceptron to predict probabilities for the 8 coarse classes and hemoglobin values. Model ensembling strategy is used in the testing step. We obtain cell level

attentions from the transformer aggregator. **(B)** Our model distinguishes acute leukemias, MPN and healthy patients with a precision of over 0.75. cAItomorph shows surprisingly high performance on MDS and plasma cell neoplasms. **(C)** cAItomorph achieves 0.97 area under the curve for malignant vs. non-malignant separation. **(D)** Predicted and measured hemoglobin values show high correlation (Pearson coefficient = 0.67). **(E)** cAItomoprh model outputs are well calibrated, matching the true probabilities. **(F)** Model performance saturates with 50% of the training data, indicating that our dataset is sufficiently large to train our model architecture. **(G)** UMAP latent space embeddings show well distinguishable patient clusters corresponding to different disease classes.

# 3. Results

### 3.1. cAItomorph identifies acute cases and myeloproliferative neoplasms with high accuracy

The overall accuracy of the model is 0.72 (Figure 2B). We achieve 0.99 sensitivity for the malignant vs. non-malignant prediction, and specificity of 0.88 (AUROC of 0.97, Figure 2C, D). Some leukemia types, including acute leukemia (0.74) and MPN (0.85) achieve a high sensitivity (0.77). Surprisingly, cAItomorph achieves high sensitivity on plasma cell neoplasms (0.70) despite the fact that these are typically not detectable through cytomorphology but relatively lower precision (0.60). Our model achieves a reasonably high performance for MDS, which is also typically not diagnosed based on peripheral blood cytomorphology alone. For lymphoma, the sensitivity is as expected low (0.40), since the disease is only diagnosable for a fraction of cases where the number of lymphocytes are significantly increased or malignant cells infiltrate into peripheral blood. Similarly, low sensitivity was achieved for MDS/MPN (0.50) and reactive changes (0.45). These disease classes are notoriously difficult to diagnose from peripheral blood. However, all sensitivities were at least 3-fold higher than random guessing (0.125, Figure 2B) suggesting that cAItomorph is able to detect morphological features that a human cytologist struggles with.

In many machine learning architectures, including ours, the final layer produces logits that are subsequently transformed by the softmax function into output values. These values do not necessarily represent actual probabilities, which is particularly critical in medical applications. To ensure interpretability, we plot calibration reliability diagrams[32,33]. cAItomorph ensemble model output probabilities match to true probabilities with expected calibration error (ECE) of 2.4% (Figure 2F). After ensuring that model probabilities were calibrated, we assessed the top-2 prediction performance (Supplementary Figure 1), where a prediction is considered correct if the correct label is among the two classes with the highest probabilities. cAItomorph achieves an average top-2 accuracy of 0.87. Notably, MDS, lymphoma, MDS/MPN, and reactive changes reach top-2 accuracies of 0.84, 0.74, 0.68, and 0.76 (Supplementary Figure 1A). For cases not detected at top-1 but identified at top-2, we analyze the differences between the highest and the second-highest predictions, expressed as delta probability (Supplementary Figure 1B). For MDS, 5 cases are predicted correctly at top-2, and 3 out of 5 have a delta probability of less than 0.25. This indicates that although the model may not predict the correct class initially, it is not highly confident in its top-1 prediction in those cases. For lymphoma, 19 cases are predicted correctly at top-2; 6 out of 19 have a delta probability below 0.25, while 13 out of 19 are below 0.5. For reactive changes, the model correctly identifies 13 cases at top-2, with 5 of them having a delta probability below 0.25. Overall, this suggests that the model's uncertainty in such cases is reflected by small margins between the top predictions.

Acute leukemias usually present immature leukocytes in the blood and are easier to identify compared to other classes in the dataset. cAItomorph has 0.70 sensitivity on acute myeloid leukemia (AML) cases, assigning 9/37 cases to MDS (Figure 3A). To analyse the misclassified cases, we plot the myeloblast ratio in the peripheral blood and model output probability for the disease (Figure 3B). cAItomorph acute leukemia output probability correlates with myeloblast ratio in the blood. All misclassified cases have lower myeloblast ratio and are assigned as MDS.

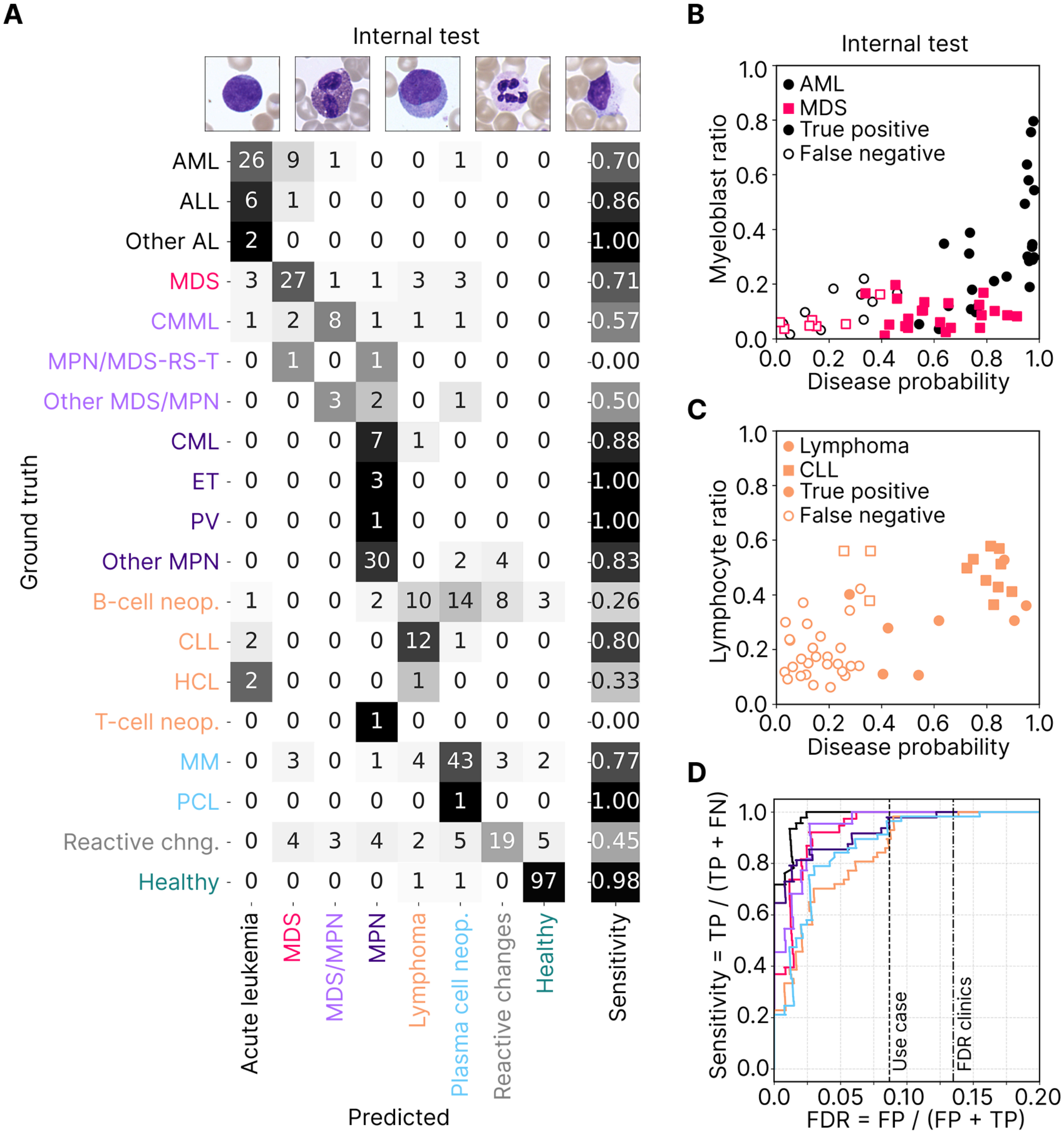

**Figure 3. Confusion matrix of detailed classes reveals high sensitivity in subtypes with visible aberrations in peripheral blood. (A)** cAItomorph achieves high sensitivities in subtypes with visible aberrations or high cellular counts, such as AML, ALL, CLL, ET, and CML. cAItomorph also achieves high sensitivity on multiple myeloma cases although it is not typically diagnosed from morphology. As expected, performance was lower in disease classes not diagnosed from peripheral blood, such as lymphoma (excluding CLL). The model almost perfectly identified healthy donors, but, as expected, it struggles with patients exhibiting reactive changes. **(B)** cAItomorph acute leukemia prediction probability correlates with myeloblast ratio in the blood. **(C)** cAItomorph lymphoma prediction probability correlates with lymphocyte ratio in the blood. The cases with high lymphocyte ratio belong to the CLL subtype. **(D)** Sensitivity versus false discovery rate for malignancy detection, stratified by disease. cAItomorph reduces the false discovery rate from 0.135 to 0.087 without missing any acute leukemia cases, based solely on peripheral blood cytomorphology.

MPN is a clonal disorder of hematopoietic stem cells characterized by proliferation of one or more myeloid lineages, resulting in mature cell overproduction. Our model correctly classifies almost all cases with essential thrombocytosis (ET), polycythemia vera (PV), chronic myeloid leukemia (CML) and shows 0.83 sensitivity in other MPN cases. In contrast, MDS is characterized by varying degrees of single or multiple cytopenias[4] and dysplasia. cAItomorph effectively distinguishes these differences from peripheral blood cell images, misclassifying only 1 of 38 MDS cases as MPN (Figure 2B). MDS/MPN is a group of myeloid neoplastic diseases with overlapping clinical and pathologic features of both MDS and MPN, described as 'cytopenia' together with 'cytosis' in the WHO 2022 classification[4]. Thus, there was an expected confusion in MDS/MPN cases, 50% of them are predicted correctly; 13.6% are predicted as MDS, and 18.2% were predicted as MPN.

There is also considerable confusion between lymphoma and plasma cell neoplasm cases. Specifically, 26% of lymphoma patients are misclassified as plasma cell neoplasm. Examination of the detailed confusion matrix (Figure 3A) reveals that these are primarily B-cell lymphoma cases other than chronic lymphocytic leukemia (CLL). cAItomorph achieves a sensitivity of 0.80 in identifying CLL cases but only 0.26 in other B-cell neoplasms. Given that plasma cell neoplasms and a fraction of lymphomas do not present in peripheral blood, the model tends to assign such cases to the plasma cell neoplasm class when there is no clear lymphocytic infiltration, as seen in CLL. We conduct a detailed analysis of lymphoma cases to understand the reasons for this diagnostic confusion. By plotting lymphoma prediction probability against lymphocyte ratio in blood, we find that the model achieves high accuracy when lymphocyte counts are elevated (Figure 3C). Notably, most of these cases belong to the CLL subtype.

Overall, cAItomorph achieves high accuracy in diseases with visible cytological aberrations, such as acute leukemias, or in cases characterized by increased counts of specific cell types, such as CLL and MPN. It nearly perfectly identifies healthy donors (sensitivity 0.98). Additionally, we confirm that our dataset is sufficiently large to train a reliable model: performance saturates when using more than 50% of the original training set (Figure 2G).

A UMAP of patient embeddings show clusters corresponding to coarse disease classes, with a clear split between myeloid and lymphoid branches (Figure 2H). The myeloid branch extends from acute leukemias through MDS to MPN, while the lymphoid branch extends from lymphoma through reactive changes to plasma cell neoplasms. Healthy individuals are well separated from malignancies.

Detecting malignancy in peripheral blood is crucial whether to perform bone marrow aspiration, a painful and invasive procedure. By summing the probabilities of all malignant classes, we provide malignancy probability, which can be used to guide further testing. We evaluate our model's sensitivity and false discovery rate (FDR) in terms of binary malignancy prediction. FDR measures the proportion of patients unnecessarily recommended for bone marrow aspiration, with an ideal value of 0. We calculated the FDR for German clinics as 13.5%, based on the number of suspected patients who underwent bone marrow aspiration but were subsequently found to have reactive changes (Figure 2B, 3D). It is important to note that we lack sensitivity data, as information on undiagnosed or later-diagnosed leukemia cases is not available. By adjusting the model's malignancy sensitivity threshold, we can balance between detecting every leukemia case and avoiding unnecessary bone marrow aspirations. In our use scenario (0.5 malignancy threshold, Figure 3D), we reduce the FDR from 13.5% to 8.7% (a 35.5% relative reduction) based solely on peripheral blood morphology without missing any acute leukemia cases.

To assess the generalizability, we evaluate our model on two external datasets. The AML-Hehr dataset includes four genetic subtypes of AML and healthy stem cell donors, while the APL-AML dataset comprises only AML patients, grouped into acute promyelocytic leukemia (APL) and other AML subtypes. In the AML-Hehr dataset, our model misclassifies only 8 out of 129 AML patients, achieving a sensitivity of 0.94 for acute cases and an AUROC of 0.99 in distinguishing malignant from non-malignant samples (Figure 4A). In the APL-AML dataset, the model shows a sensitivity of 0.76 for detecting acute cases and correctly identifies all as malignant (Figure 4B). Notably, the APL-AML dataset is completely out-of-domain, having a distinct staining background. We visualize external patient embeddings onto the same UMAP fitted on internal data.

Remarkably, external cases aligned closely with internal test samples, forming coherent clusters consistent with their disease types (Figure 4C).

We further evaluate our model on an extended test set including patients with unclear/rare diagnoses, double diagnoses, and MGUS cases previously excluded (Figure 4D). For this analysis, we group the diagnoses into broader categories and introduce a new 'other' class (Figure 4D). Our model is able to classify borderline cases in one of the suspected classes. For instance, 5/11 assigned as acute leukemia and 4/11 assigned as MDS in MDS-AML borderline cases. MPN in blast crises are assigned either in acute leukemia (3/4) or MPN (1/4). More surprisingly, cAItomorph classifies 38% of MGUS cases as plasma cell neoplasms, which is a precursor condition of multiple myeloma and typically not determined from morphology.

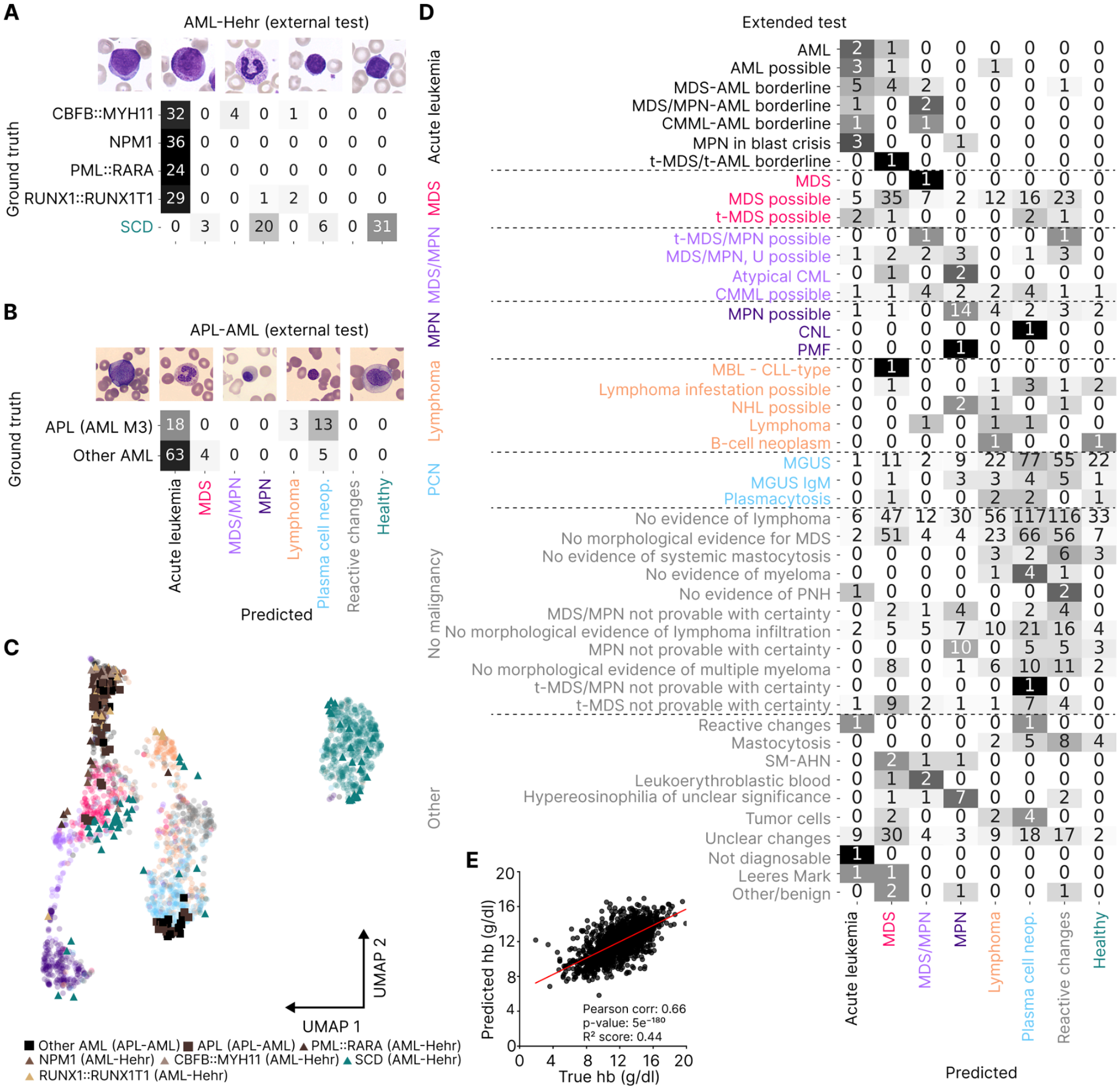

**Figure 4.** cAItomorph identifies acute leukemia cases with high sensitivity on external datasets and detects borderline cases. **(A)** Model performance on the AML-Hehr dataset. Ground-truth labels include four genetic subtypes of AML and stem cell donors (SCD). cAItomorph identifies all acute leukemia cases with high sensitivity and precision. **(B)** Confusion matrix for predictions on the APL-AML dataset. **(C)** External patient embeddings align consistently with internal data according to their corresponding labels in the UMAP space. **(D)** Confusion matrix for the extended test set, which includes borderline, hematologically suspicious but non-malignant, rare, and undiagnosed cases. cAItomorph successfully identifies probable AML, MPN, and borderline AML cases. **(E)** Predicted hemoglobin correlates with measured values in the extended test set.

### 3.2. cAItomorph highlights clinically relevant cells

The ultimate goal of our AI model is to support hematologists in clinical practice. Our model provides eight-class hematological disease classification, malignancy probability estimation, and cell-level attention maps. The cell-level attention highlights the most diagnostically relevant cells, ensuring that the model's decisions are based on the correct cell types.

We present two patients from the internal test set and two from external test sets as illustrative examples (Figure 5). For each patient, ground-truth metadata and the predicted hemoglobin value are provided. Adjacent to the metadata, the eight-class hematological diagnosis and malignancy probability are displayed. In addition to patient-level predictions, we generate single-cell predictions by passing each white blood cell image individually through cAItomorph to better understand correlations between cell and disease types. In the figure, cells are colored according to their single-cell predictions and sorted based on the attention scores provided by the model for explainability.

cAItomorph successfully assigns high attention to myeloblasts when classifying an acute leukemia case, while assigning minimal attention to typical lymphocytes (Figure 5A). In a patient with myeloproliferative neoplasm (MPN), the model highlights giant platelets that can hint to the diagnosis of MPN (Figure 5B). In a patient with *CBFB::MYH11* fusion from the external AML-Hehr dataset, the model identifies myeloblasts and monocytic cells, consistent with their association with myelomonocytic leukemias (Figure 5C). In an APL patient from the APL-AML dataset, cAItomorph highlights promyelocytes and myeloblasts in the peripheral blood—cells that are diagnostically defining for acute promyelocytic leukemia (Figure 5D).

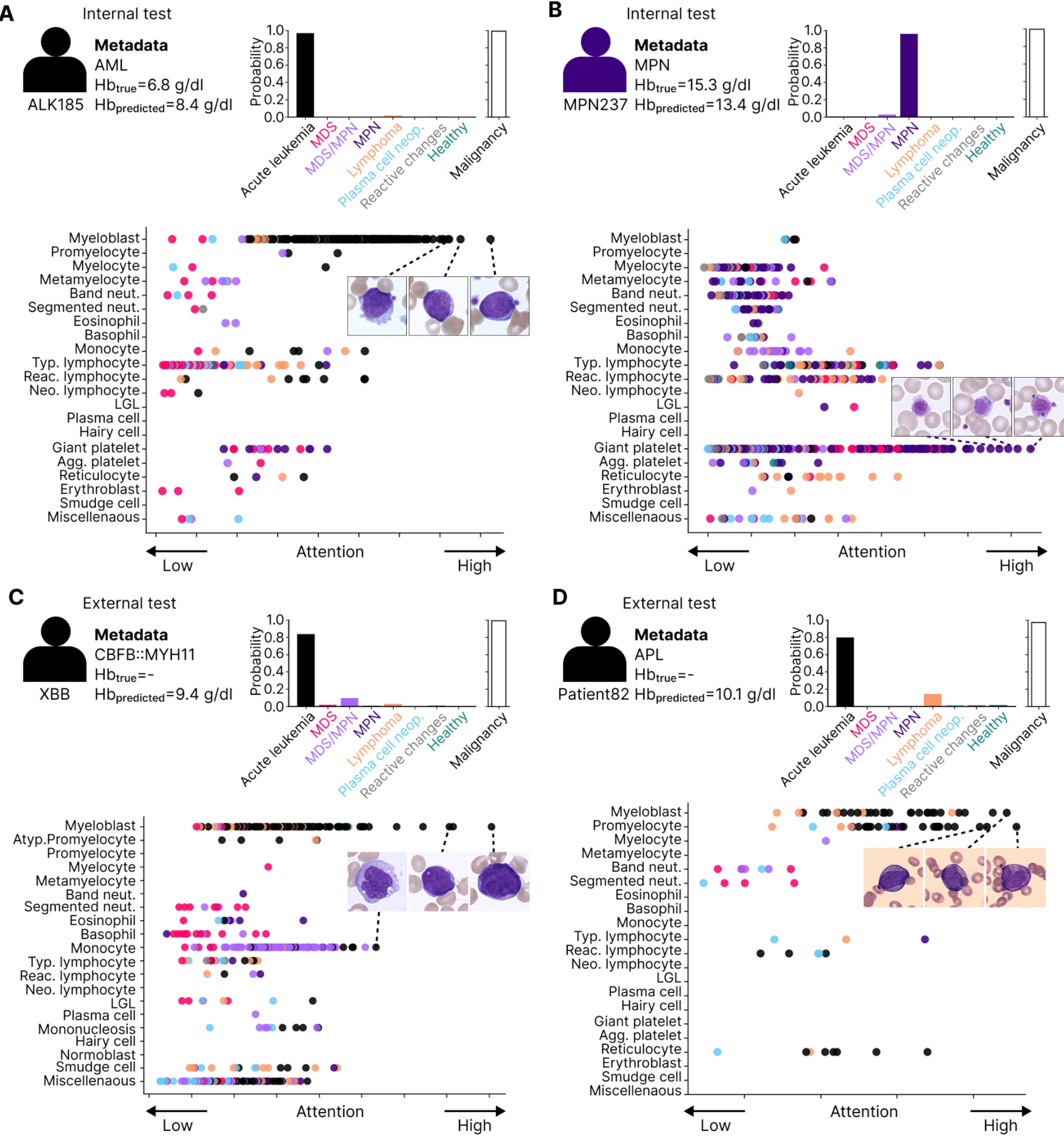

**Figure 5. cAItomorph detects clinically relevant cells for diagnosis.** Four exemplary patients are presented: two from the internal test set, one from the AML-Hehr dataset, and one from the APL-AML dataset. For each patient, we provide metadata, predicted hemoglobin, eight-class hematological disease probabilities, malignancy probability, and attention distribution over cells. Individual cells are passed through cAItomorph to better understand their relation to diseases. Cell dots are colored based on model predictions. **(A)** An AML patient from the internal test set, confidently predicted as acute leukemia by cAItomorph. Myeloblasts receive the highest attention. **(B)** An MPN patient from the internal test set, predicted to be an MPN and malignant. cAItomorph assigns high attention to giant thrombocytes, which supports the diagnosis of MPN. Predicted hemoglobin values correlate with the magnitude of the actual measured values, although they do not match exactly. **(C)** An AML case with a CBFB::MYH11 fusion mutation from the AML-Hehr dataset. cAItomorph diagnoses the case with high confidence while assigning high attention to myeloblasts and monocytic cells. **(D)** An APL patient from the APL-AML dataset. Promyelocytes and myeloblasts receive the highest attention, as expected.

## 4. Discussion

We present a patient-level hematology model, cAItomorph, that is trained on real-world peripheral blood cytomorphology data to predict bone marrow malignancies. To develop the model, we collected the first real-world peripheral blood cytomorphology data set with over 6000 patients and 3.2 million single cell images from a wide range of hematological diseases. Using the hematology foundation model, we predict six common coarse hematological malignancies including acute leukemias, myelodysplastic changes, myeloproliferative neoplasms, lymphomas, plasma cell neoplasms along with reactive and healthy conditions. cAItomoprh detects We also estimate hemoglobin value of a patient using only single cell images.

The standard procedure in hematological diagnostics involves a comprehensive set of tests, including cyto-/histomorphology, immunophenotyping, cytogenetics, and molecular genetics. These tests often require different tissue materials, such as bone marrow and lymph node biopsies depending on the suspected disease. Peripheral blood smear is a crucial initial test as it reflects abnormalities originating in the bone marrow. Given its wide availability and low cost, we were motivated to develop a tool that can initially identify abnormalities in the blood and guide subsequent diagnostic testing.

Over the past few years, there has been a stepwise evolution in the application of computer vision to hematological malignancy detection. Initial efforts focused on cell identification, segmentation, and classification in bone marrow and peripheral blood smears[11–14,16,17]. Subsequent research shifted toward detecting abnormal cells leading to disease detection, such as achieving human-level recognition of blast cells in AML using convolutional neural networks[5]. Studies on patient level diagnosis are more limited. Some studies provided patient-level diagnosis based on single-cell or patch information[19,22,26]; however, training such models requires extensive annotation. Recent studies[18,24,34], including ours, translate multiple instance learning—a pipeline widely used in computational pathology—into the hematology field, eliminating the need for single-cell annotation. Our training set size and disease range are larger than those of previous studies, which primarily focused on a limited number of diseases[18,19,24,26,34,35].

Ablation analysis reveals that using transformer based architecture is superior to both widely used ABMIL[36] and variants of attention based aggregators[24], regardless of the number of transformer layers used (Supplementary Table 1, 2). Increasing the number of transformer layers almost linearly enhances model performance (Supplementary Table 2), consistent with the known phenomenon[37,38]. We also ablate the training set size using the best-performing model architecture and observe that similar performance can be achieved with less data. Promisingly, performance can be further improved by enhancing single-cell representation quality, increasing aggregator model size, and incorporating additional data from diverse sources.

The output probabilities of cAItomorph correspond well to the true probabilities. This is important to ensure that the model is not overconfident in its diagnoses. We sum the probabilities of malignant classes to decide whether a patient requires further testing, such as bone marrow aspiration. In our internal test data, we demonstrate a scenario that reduces the number of bone marrow aspirations while not missing any acute leukemia cases, with less than 2% missed in most disease categories. Notably, this is based solely on peripheral blood cytomorphology and can be used as a screening tool in settings where hematologists are scarce, such as underserved regions and general practitioner offices. Moreover, our model can support clinical decision-making by providing cell-level explainability, highlighting which cells are most important for the diagnosis—an aspect that is also legally relevant[39].

Our study is not without limitations. First, our real-world dataset is unbalanced with respect to class distributions, making the detection of rare diseases more challenging. Second, the healthy cohort in our training and test data consists exclusively of stem cell donors. Stem cell donors are typically younger individuals who are medically fit and free from serious cardiovascular disease, autoimmune disorders, chronic infections, or other long-standing illnesses. It is well established that chronic conditions can alter peripheral blood cytomorphology, even in the absence of direct hematologic disorders. Consequently, our model may exhibit a bias toward assigning a malignant label when a person is not completely healthy, as observed in the extended test set. A prospective study design is needed for further testing.

This work aims at developing and testing an AI algorithm in real-world settings, opening possibilities for future development and applications. We envision cAItomorph being integrated into the hematology workflow and serving as a cornerstone for future multimodal and agentic AI systems.

## Ethics statement

All experiments are conducted in accordance with the Declaration of Helsinki. The retrospective analysis of images used in study received approval from Ludwig Maximilian University of Munich ethics committee.

## Acknowledgements

C.M. received funding from the European Research Council under the European Union's Horizon 2020 Research and Innovation Programme (grant agreement 866411 & 101113551). We acknowledge support from the Hightech Agenda Bayern.

## Code and data availability

The test set used in this study, along with the code and model weights, will be made publicly available upon publication. Our code and model weights are available at https://github.com/marrlab/cAItomorph.

External test sets are available under the corresponding links:

AML-Hehr: https://www.cancerimagingarchive.net/collection/aml-cytomorphology_mll_helmholtz/
APL-AML: https://www.kaggle.com/datasets/eugeneshenderov/acute-promyelocytic-leukemia-apl

## Authorship

**Contributions:** C.P., C.M. and I.K., conceptualized the study idea. C.P., P.L. and I.K. curated the data. M.F.D., I.K. and F.O. trained the models, conducted formal analysis, interpreted and visualized the results. I.K., M.F.D., M.H. and C.M. drafted the original manuscript. C.M. and C.P. supervised the study. K.S. and M.H. provided clinical expertise. K.S., A.S., M.H., P.L., and C.P. critically revised the manuscript. All authors had full access to the data, made a review of the paper, and approved the final submission.

**Conflict of interest:** The authors declare no competing financial interests.

**Correspondence:** Carsten Marr, Institute of AI for Health, Helmholtz Zentrum München—German Research Center for Environmental Health, Neuherberg, Germany; e-mail: carsten.marr@helmholtz-munich.de.

## Supplementary Figures & Tables

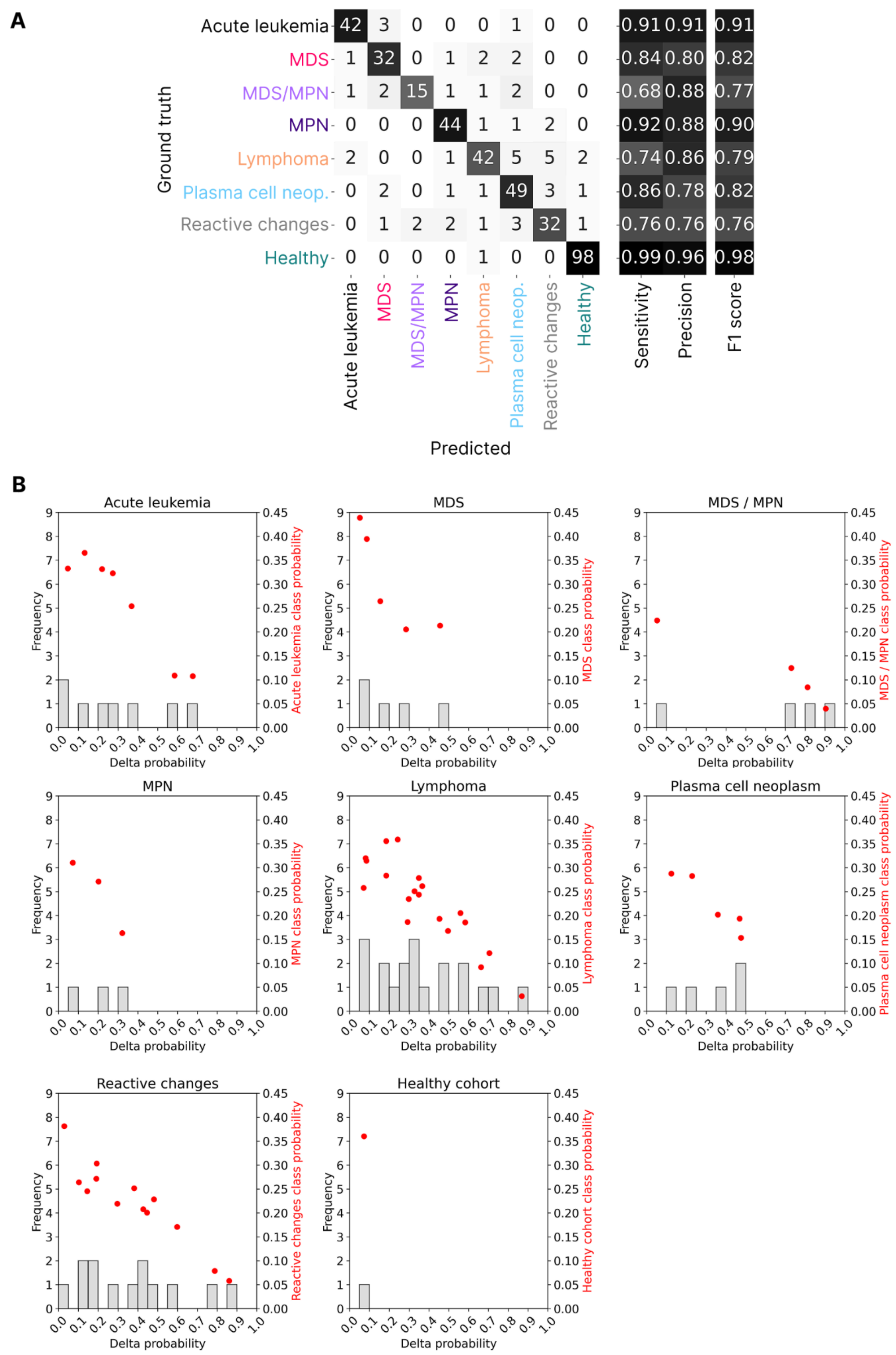

**Supplementary Figure 1. Model performance improves at top-2 prediction, especially for lymphoma, MDS/MPN, and reactive changes. (A)** The model generates a logits vector, which is converted into probabilities using the softmax function. We analyze the model's performance for the top-2 prediction, where a prediction is considered correct if the correct label is among the two classes with the highest probabilities.

**(B)** For cases not detected at top-1 but at top-2, we calculate the probability differences between the highest and the second highest predictions, expressed as delta probability. Gray bars represent the number of cases within each delta probability interval, with each interval corresponding to a 0.05 probability difference. Red dots indicate individual instances and their corresponding second-highest probabilities at certain delta probability.

**Supplementary Table 1.** Comparison of different aggregator models.

| | Balanced accuracy | Weighted F1 |
|---|---|---|
| Transformer | **0.64±0.03** | **0.68±0.03** |
| WBCMIL | 0.60±0.02 | 0.65±0.01 |
| ABMIL | 0.52±0.01 | 0.57±0.01 |
| Mean | 0.60±0.02 | 0.65±0.02 |

**Supplementary Table 2.** Ablation on different pooling strategy in transformer model and number of transformer layers. * denotes the architecture used in this study

| Pooling | Depth | Balanced accuracy | Weighted F1 |
|---|---|---|---|
| CLS pooling | | | |
| CLS | 2 | 0.61±0.02 | 0.66±0.03 |
| CLS | 4 | 0.61±0.01 | 0.66±0.01 |
| CLS | 6 | 0.62±0.02 | 0.67±0.03 |
| CLS | 8 | 0.63±0.02 | 0.68±0.02 |
| *CLS | 10 | 0.64±0.03 | 0.68±0.03 |
| CLS | 12 | 0.63±0.02 | 0.67±0.02 |
| Mean pooling | | | |
| Mean | 2 | 0.63±0.02 | 0.68±0.02 |
| Mean | 4 | 0.64±0.02 | 0.68±0.02 |
| Mean | 6 | 0.63±0.02 | 0.68±0.01 |
| Mean | 8 | 0.64±0.01 | 0.68±0.02 |
| Mean | 10 | 0.62±0.02 | 0.67±0.02 |
| Mean | 12 | 0.63±0.01 | 0.68±0.01 |